\documentclass[twocolumn, 10pt, pra, aps, showpacs, reprint, amsmath, amssymb, superscriptaddress, nofootinbib]{revtex4-1}

\usepackage{graphicx, bm}

\newlength\figurewidth
\newcommand{\rem}[1]{}

\newcommand{\imagc}[1]{\text{Im}\,#1}
\newcommand{\real}[1]{\text{Re}(#1)}

\newcommand{\limacon}{{lima\c con}}
\newcommand{\lopt}{\ell_\mathrm{opt}}
\newcommand{\lgeo}{\ell_\mathrm{geo}}

\begin{document}

\title{Unidirectional light emission from low-index polymer microlasers}
\author{M. Schermer}
\affiliation{Institut f{\"u}r Theoretische Physik, Otto-von-Guericke-Universit{\"a}t Magdeburg, Postfach 4120, D-39016 Magdeburg, Germany}
\author{S. Bittner}
\author{G. Singh}
\affiliation{Laboratoire de Photonique Quantique et Mol{\'e}culaire, CNRS UMR 8537, Institut d'Alembert FR 3242, {\'E}cole Normale Sup{\'e}rieure de Cachan, 61 avenue du Pr\'esident Wilson, F-94235 Cachan, France}
\author{C. Ulysse}
\affiliation{Laboratoire de Photonique et Nanostructures, CNRS UPR20, Route de Nozay, F-91460 Marcoussis, France}
\author{M. Lebental}
\email{lebental@lpqm.ens-cachan.fr}
\affiliation{Laboratoire de Photonique Quantique et Mol{\'e}culaire, CNRS UMR 8537, Institut d'Alembert FR 3242, {\'E}cole Normale Sup{\'e}rieure de Cachan, 61 avenue du Pr\'esident Wilson, F-94235 Cachan, France}
\author{J. Wiersig}
\email{jan.wiersig@ovgu.de}
\affiliation{Institut f{\"u}r Theoretische Physik, Otto-von-Guericke-Universit{\"a}t Magdeburg, Postfach 4120, D-39016 Magdeburg, Germany}
\date{ \today}

\begin{abstract}
We report on experiments with deformed polymer microlasers that have a low refractive index and exhibit unidirectional light emission. We demonstrate that the highly directional emission is due to transport of light rays along the unstable manifold of the chaotic saddle in phase space. Experiments, ray-tracing simulations, and mode calculations show very good agreement.
\end{abstract}
\maketitle

The physics of optical microcavities is a topical research field for more than one decade~\cite{Vahala03}. Microcavities~\cite{MLSGPL92,MSJPLHKH07} confine photons for a long time $\tau$ due to total internal reflection at the boundary of the cavity. These so-called whispering-gallery modes (WGMs) have high quality factors $Q = \omega\tau$, where $\omega$ is the resonance frequency. The in-plane light emission from an ideal circular-shaped microlaser is isotropic due to the rotational symmetry. To overcome this disadvantage microcavities with deformed boundaries have been fabricated leading to significantly improved emission patterns~\cite{LSMGPL93, ND97, GCNNSFSC98, SRTCS04, Cao2015}. Even unidirectional emission is possible, which has been demonstrated for several shapes, e.g., the spiral~\cite{CTSCKJ03,HK09,HKB09}, cavities with holes~\cite{WH06, Djellali2009}, the {\limacon}~\cite{WH08,SCL08,YWD09,WYD09,SHH09,YKK09,AHE12}, the circle with a point scatterer~\cite{DMS09}, and the notched ellipse~\cite{BBS06,WYY10}. 

The ray dynamics inside a deformed microcavity is (partially) chaotic, i.e., neighboring ray trajectories deviate from each other exponentially fast. Because of this, deformed microdisks have attracted attention as models for studying ray-wave correspondence in open systems~\cite{ND97}. This is analog to the study of quantum-classical correspondence in the field of quantum chaos~\cite{Stoeckmann00}. In open chaotic systems the long-time behavior of trajectories is governed by the chaotic saddle (or chaotic repeller for noninvertible dynamical systems) and its unstable manifold \cite{LichLieb92,LT10}. The chaotic saddle is the set of points in phase space that never visits the leaky region both in forward and backward time evolution. The unstable manifold of a chaotic saddle is the set of points that converges to the saddle in backward time evolution. This unstable manifold therefore describes how trajectories, after a transient time, escape from the open chaotic system. It has been shown both theoretically and experimentally that this kind of unstable manifold in chaotic microcavities determines the far-field pattern of all high-$Q$ modes~\cite{SRTCS04,LYMLASLK07}. As a consequence, the high-$Q$ modes in such a cavity have very similar far-field patterns. This useful property has been termed universal far-field pattern~\cite{LYMLASLK07}. Based on this concept it was predicted that the light emission from high-$Q$ modes in a {\limacon} cavity with refractive index $n$ between 2.7 and 3.9 is universal and unidirectional~\cite{WH08}. This was confirmed experimentally by a number of groups~\cite{SCL08,YWD09,WYD09,SHH09,YKK09,AHE12}.

Microcavities made from materials with low refractive index $n \leq 2$ like polymers are interesting due to their cheap and easy fabrication. For this low-index regime, the ``face'' cavity has been proposed~\cite{ZouJSTQE13}. Unfortunately, this cavity is not fully chaotic and therefore the universality and directionality of light emission is spoiled by islands of regular motion in phase space. The notched ellipse can work for a rather broad regime of refractive indices by adapting the eccentricity of the ellipse~\cite{WYY10}. In the low-index regime, however, the notch does not function as an efficient scatterer needed for directing the light~\cite{Unterhinninghofen11}. Deformed silica microtoroids can also exhibit directional emission but at the cost of a complicated fabrication process \cite{Jiang2012}.

\begin{figure}[tb]
\includegraphics[width = 8 cm]{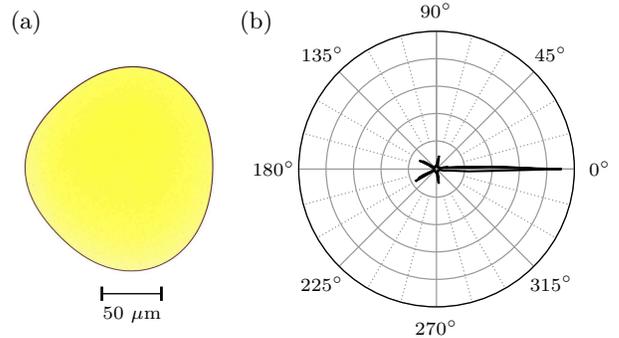}
\caption{(a) Top view of a shortegg microlaser with $R=80$ $\mu$m in real colors with an optical microscope. (b) Measured far-field intensity pattern of a polymer shortegg microlaser in a polar plot as a function of the azimuthal angle~$\phi$.}
\label{fig:photo+diagramme}
\end{figure}

It is the aim of the present letter to fill this low-index gap by introducing a cavity shape called ``shortegg'' that is shown in Fig.~\ref{fig:photo+diagramme}(a). For $1.5 \leq n \leq 1.8$, its emission is strongly concentrated in a single direction [see Fig.~\ref{fig:photo+diagramme}(b)], the far-field pattern is universal, and the quality factors are reasonably high. Numerical simulations and experimental data show good agreement.

The boundary of the shortegg cavity is given in polar coordinates by
\begin{equation} \label{eq:shortegg}
\rho(\phi) = R [1 + \varepsilon_1 \cos(\phi) + \varepsilon_2 \cos(2\phi) + \varepsilon_3 \cos(3\phi)]
\end{equation}
with mean radius $R$ and deformation parameters $\varepsilon_1 = 0.16$, $\varepsilon_2 = -0.022$, and $\varepsilon_3 = -0.05$. The ray dynamics inside microcavities is best described in a two-dimensional phase space representation, the so-called Poincar\'e surface of section. Whenever the ray hits the cavity's boundary, its position $s$ in terms of the arclength coordinate along the circumference and its tangential momentum $\sin(\chi)$ are recorded. Here, $\chi$ is the angle of incidence measured from the surface normal. An angle $\chi < 0$ ($\chi > 0$) indicates clockwise (counterclockwise) propagation direction. When $|\chi|$ is larger than the critical angle for total internal reflection, $\chi_c = \arcsin{(1/n)}$, the ray is completely reflected. In the leaky region of phase space $|\chi| < \chi_c$ the ray is only partially reflected according to Fresnel's law.

\begin{figure}[tb]
\includegraphics[width = 7.0 cm]{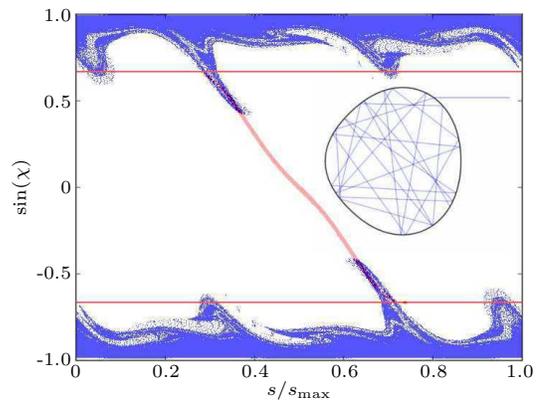}
\caption{Unstable manifold (blue points) of the chaotic saddle in the shortegg cavity for TE polarization. The red horizontal lines mark the border of the leaky region $|\chi| < \chi_c$. The solid red curve is the set of points that leave the cavity towards the far-field angle $0^\circ$. Inset: typical trajectory from the unstable manifold.}
\label{fig:unstablemanifold}
\end{figure}

The ray dynamics inside the shortegg is chaotic except for small islands of regular motion in the leaky region (not shown). Figure~\ref{fig:unstablemanifold} depicts the unstable manifold of the chaotic saddle for the case of transverse electric (TE) polarization (i.e., electric field parallel to the plane of the cavity) with effective refractive index $n = 1.5$ according to the numerical scheme discussed in Ref.~\cite{WH08}. It looks similar for transverse magnetic (TM) polarization (i.e., magnetic field parallel to the plane). It should be noted that flat dye-based microlasers predominantly exhibit TE polarized modes \cite{Gozhyk2012}. As in the situation of the {\limacon} cavity~\cite{WH08}, the unstable manifold has only two significant overlaps with the leaky region, one at $s/s_\mathrm{max} \approx 0.35$ and one related by symmetry at $s/s_\mathrm{max} \approx 0.65$. However, in contrast to the {\limacon} cavity these two overlap regions are rather elongated and it is therefore not clear \textit{a priori} how this can lead to unidirectional emission. In fact, we have optimized the boundary of the cavity such that the arms of the unstable manifold follow precisely the curve of points that are emitted to the far-field angle $0^\circ$ (red curve in Fig.~\ref{fig:unstablemanifold}). Due to this fact, a highly directed emission can be expected.

\begin{figure}[tb]
\includegraphics[width = 7.0 cm]{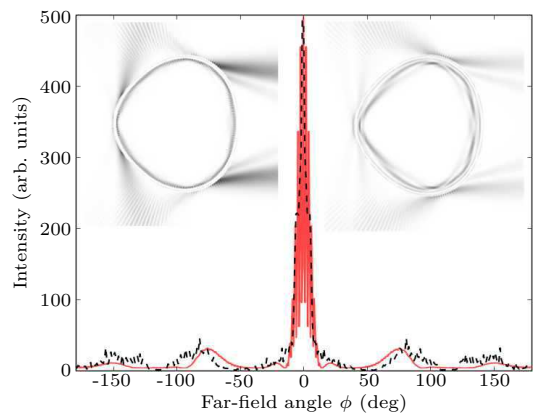}
\caption{Far-field intensity pattern from ray-tracing simulations (dashed black curve) and mode calculations (solid red curve) for TE polarization. Left inset: near-field intensity pattern of the corresponding WGM with radial mode number 1 and dimensionless frequency $\real{\Omega} \approx 89.41$, and $Q \approx 25\,800$. Right inset: near-field intensity pattern of a whispering-gallery-like mode with radial mode number 3, $\real{\Omega} \approx 89.28$, and $Q \approx 8\,000$. White corresponds to minimal and black to maximal intensity. The intensity outside of the cavity is enhanced by a factor of 30 (left) and 10 (right) to highlight the light emission.}
\label{fig:ffpwave}
\end{figure}

The far-field intensity pattern resulting from the ray-tracing simulations is shown as dashed black curve in Fig.~\ref{fig:ffpwave}. As predicted by the structure of the unstable manifold, the emission is strongly peaked around the far-field angle $\phi = 0^\circ$ with considerably smaller side peaks at $\pm 80^\circ$ and $\pm 150^\circ$. The beam divergence is as small as $\pm 3^\circ$, which is much smaller than the beam divergence from the {\limacon} cavity~\cite{WH08} and comparable to the beam divergence from the notched ellipse~\cite{WYY10} in the high-index regime. The small oscillations of the far-field intensity pattern in Fig.~\ref{fig:ffpwave} are due to the nontrivial fine structure of the unstable manifold which is not visible on the scale shown in Fig.~\ref{fig:unstablemanifold}. Due to the small value of $n$ the emission for TM polarization is similar with additional small side peaks at $\pm 120^\circ$ (not shown). Extensive numerical investigations reveal that the unidirectional emission persists for $1.5 \leq n \leq 1.8$ but not for higher refractive indices. Moreover, the far-field pattern is robust with respect to small variations of the deformation parameters $\varepsilon_1$, $\varepsilon_2$, and $\varepsilon_3$. However, the directionality is considerably spoiled for $\varepsilon_3 = 0$.

Numerical simulations of Maxwell's equations in the shortegg geometry are performed within the effective refractive index approximation in two dimensions (see, e.g., Ref.~\cite{Lebental2007}). Using the boundary element method~\cite{Wiersig02b} we determine the spatial mode pattern $\psi(x,y)$ corresponding to the $z$-component of the magnetic (TE) and electric (TM) field as well as the complex resonant frequencies $\omega =ck$ where $c$ is the speed of light in vacuum and $k$ is the wave number. The real part is the conventional frequency whereas the imaginary part determines the lifetime $\tau =-1/[2\,\imagc{\omega}]$ of a given mode. For convenience, the dimensionless complex frequency $\Omega = \omega R/c = kR$ is used.

In the considered frequency regime the computed high-$Q$ modes are WGMs (two examples for $n = 1.5$ are shown as insets in Fig.~\ref{fig:ffpwave}). In all cases their far-field patterns agree well with the ray-tracing simulations, which clearly demonstrates the universality of the far-field patterns. The fact that WGMs with reasonably high $Q$s are formed in such a strongly deformed cavity with chaotic ray dynamics can be explained by the existence of partial barriers in phase space~\cite{SWC11}. Moreover, we verified that the Husimi functions \cite{HSS03} (not shown) of the high-$Q$ modes in the leaky region are well localized on the unstable manifold. This presents another indication that the emission directionality in the shortegg is due to the unstable manifold of the chaotic saddle.

Shortegg cavities were experimentally investigated using organic microlasers. The microlasers were fabricated from a $700$~nm thick layer of PMMA \footnote{Poly(methyl methacrylate) (PMMA A6 resist by Microchem)} doped with $5$~wt\% of the laser dye DCM \footnote{4-(Dicyanomethylene)-2-methyl-6-(4-dimethyl\-amino\-styryl)-4Hpyran (by Exciton)}. The dye-doped polymer with bulk refractive index $1.54$ was spin-coated on a Si wafer with a $2~\mu$m buffer layer of SiO$_2$ to avoid leakage of the cavity modes into the Si wafer. The cavities were created from the polymer layer by electron-beam lithography which ensured nanometric precision \cite{Lozenko2012}. They can be considered as two-dimensional systems with effective refractive index $n = 1.50$ (see Ref.~\cite{Lebental2007} for details of the effective refractive index calculation). The cavities were completely supported by the silica layer\cite{Lozenko2012}. An optical microscope image of a shortegg cavity is shown in Fig.~\ref{fig:photo+diagramme}(a). The results presented here were measured with a cavity with radius $R = 80~\mu$m [$\real{\Omega} \simeq 810$], and are similar to those obtained with smaller ones. Furthermore, a quadrupolar cavity was investigated for comparison. Its shape is given by Eq.~(\ref{eq:shortegg}) with $\epsilon_{1, 2}$ as for the shortegg and $\epsilon_3 = 0$. The microlasers were pumped by a pulsed, frequency-doubled Nd:YAG laser ($532$~nm, $0.5$~ns, $10$~Hz). The size of the pump spot was adjusted to cover the whole cavity homogeneously. The lasing emission was collected by a lens and analyzed with a spectrometer. The cavities could be rotated in order to measure the lasing spectra for arbitrary directions in the cavity plane and thus obtain the azimuthal far-field intensity pattern. See Ref.~\cite{Chen2014} for a detailed description of the experimental setup.

\begin{figure}[tb]
\includegraphics[width = 8.0 cm]{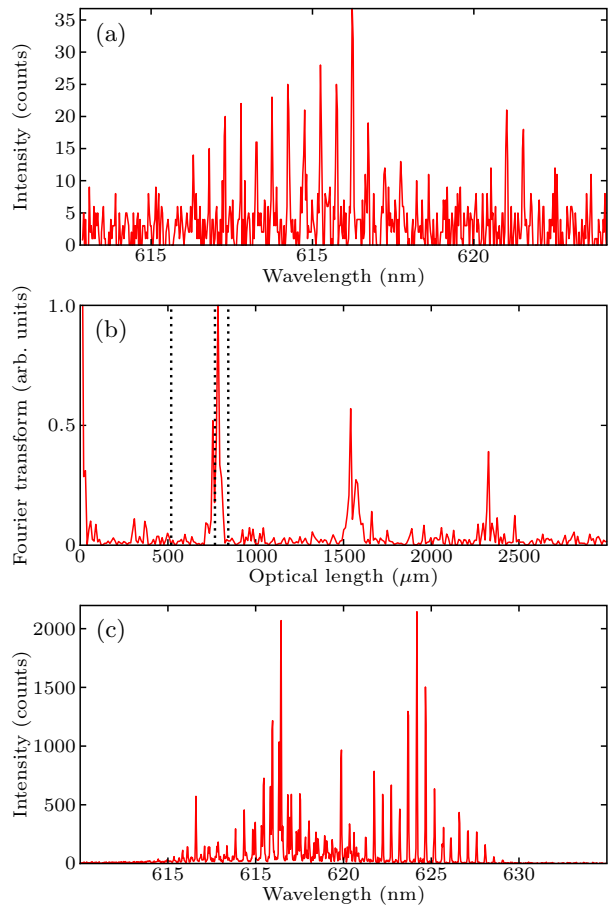}
\caption{(a) Measured lasing spectrum in the direction $\phi = 0^{\circ}$ for a shortegg microlaser with $R = 80~\mu$m just above the laser threshold. (b) Normalized Fourier transform of the spectrum in (a). The vertical dotted lines indicate the expected optical lengths of the diameter, the stable four bounce inscribed orbit ($\lgeo^\mathrm{(4b)} \simeq 5.784 \, R = 462.7~\mu$m), and the perimeter. (c) Idem (a) far above threshold.}
\label{fig:spectre}
\end{figure}

The lasing threshold intensity of the shortegg microlaser was $2.4$~MW$\cdot$cm$^{-2}$ and is about three times lower than that of the quadrupolar microlaser. This evidences that the particular shape of the shortegg leads to an enhancement of the quality factors compared to other cavities with similar shapes. The lasing emission was TE polarized.

A spectrum recorded for $\phi = 0^\circ$ just above threshold is shown in Fig.~\ref{fig:spectre}(a) and features a sequence of equidistant lasing modes. Lasing modes with large $\real{\Omega}$ are often related to specific sets of trajectories. If this is the case, then the optical length of these trajectories is inversely proportional to the free spectral range, $\lopt = 2 \pi / \Delta k$, and can be determined from the Fourier transform of the spectrum \cite{Lebental2007}. The Fourier transform of the spectrum in Fig.~\ref{fig:spectre}(a) is presented in Fig.~\ref{fig:spectre}(b). Its first two significant peaks are at $\lopt = 755.7~\mu$m and $785.5~\mu$m, and further peaks are found at multiples of these lengths. They correspond to a geometrical length of 455 $\mu$m and 473 $\mu$m, respectively, for a group refractive index of $n_g = 1.66$. These lengths are significantly longer than the length of the diameter, $\lgeo^\mathrm{(diam)} \simeq 3.912 \, R = 313~\mu$m, and hence evidence that the lasing modes are of the whispering gallery type. On the other hand they are also significantly shorter than the length of the perimeter, $\lgeo^\mathrm{(per)} \simeq 6.360 \, R = 509~\mu$m, which indicates that the observed modes are WGMs with higher radial excitation. In contrast, the spectrum of the quadrupolar microlaser (not shown) corresponds to the optical length of the diameter orbit. The shortegg cavity exhibits a larger number of resonances for higher pump intensities [see Fig.~\ref{fig:spectre}(c)], but the Fourier transform features only peaks at the same optical lengths as in Fig.~\ref{fig:spectre}(b).

The measured far-field intensity pattern in Fig.~\ref{fig:photo+diagramme}(b) shows the maximal intensity of each spectrum as a function of the azimuthal angle. It was recorded for a pump intensity two times higher than the threshold. No significant effect of bleaching was observed during the measurement. The quadrupolar microlaser exhibits a very broad far-field intensity pattern without sharp emission lobes (not shown). On the contrary, the far-field intensity pattern of the shortegg micro-lasers exhibits one dominant emission lobe at $\phi = 0^\circ$ with a divergence of about $\pm 4^\circ$, in very good agreement with the theoretical predictions. Four much smaller lobes at $\pm 80^\circ$ and $\pm 150^\circ$ are also observed. This is consistent with the numerical simulations even though the lasing modes probably have a higher radial excitation than the modes shown in Fig.~\ref{fig:ffpwave}, which once again shows the universality of the emission patterns.  

\begin{figure}[tb]
\includegraphics[width = 7 cm]{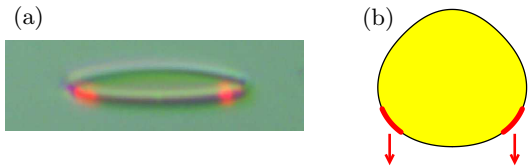}
\caption{(a) Photograph of a lasing shortegg cavity. The lasing emission appears in red (real color). The green pump beam is not visible. The direction $\phi = 0^{\circ}$ is turned towards the camera. (b) Sketch of the cavity geometry and the lasing emission corresponding to (a).}
\label{fig:camera}
\end{figure}

A photograph of the lasing cavity made by a CMOS sensor camera with a high-magnification zoom lens is presented in Fig.~\ref{fig:camera}(a). The observation direction was $\phi = 0^\circ$ and the camera had an inclination angle of $10^{\circ}$ with respect to the plane of the cavity. The photograph shows that the red lasing emission towards $\phi = 0^\circ$ originates from two small regions of the cavity boundary around $\phi = \pm50^\circ$. Figure \ref{fig:camera}(b) shows a sketch of the cavity where the emitting parts of the cavity are indicated in red. Their positions agree very well with the calculated near-field intensity patterns shown in the insets of Fig.~\ref{fig:ffpwave}.

We have demonstrated both theoretically and experimentally that low-index polymer microlasers with the so-called shortegg shape exhibit lasing modes with highly directional emission. The lasing modes are based on the unstable manifold of the chaotic saddle which leads to a universal (i.e., mode-independent) far-field pattern of the high-Q modes. The theoretical predictions of near- and far-field intensity patterns showed excellent agreement with the experimental findings. 

\vspace{2 mm}
Fruitful discussions with J.~Zyss, J.-B.~Shim and M.~Kraft are acknowledged. M.~S. thanks A.~Ebersp{\"a}cher for providing his computer code package. S.~B.\ gratefully acknowledges funding from the European Union Seventh Framework Programme (FP7/2007-2013) under Grant No.\ 246.556.10. This work was supported by a public grant from the Laboratoire d'Excellence Physics Atom Light Matter (LabEx PALM) overseen by the French National Research Agency (ANR) as part of the Investissements d'Avenir program (reference: ANR-10-LABX-0039).

\end{document}